\documentclass[aps,prl,preprint,superscriptaddress, 9pt]{revtex4-1}

\usepackage{graphicx}
\usepackage{epsfig}
\usepackage{placeins}
\usepackage{amsmath}
\usepackage{nicefrac}
\usepackage{lineno}

\begin{document}

\title{Transition of a 2D spin mode to a helical state by lateral confinement}
\author{P.~Altmann}
\affiliation{IBM Research--Zurich, S\"aumerstrasse 4, 8803 R\"uschlikon, Switzerland}

\author{M.~Kohda}
\affiliation{Department of Materials Science, Tohoku University, 6-6-02 Aramaki-Aza Aoba, Aoba-ku, Sendai 980-8579, Japan}

\author{C.~Reichl}
\affiliation{Solid State Physics Laboratory, ETH Zurich, 8093 Zurich, Switzerland}

\author{W.~Wegscheider}
\affiliation{Solid State Physics Laboratory, ETH Zurich, 8093 Zurich, Switzerland}

\author{G.~Salis}
\email{gsa@zurich.ibm.com}
\affiliation{IBM Research--Zurich, S\"aumerstrasse 4, 8803 R\"uschlikon, Switzerland}

\maketitle

{\bfseries
Spin-orbit interaction (SOI) leads to spin precession about a momentum-dependent spin-orbit field. In a diffusive two-dimensional (2D) electron gas, the spin orientation at a given spatial position depends on which trajectory the electron travels to that position. In the transition to a 1D system with increasing lateral confinement, the spin orientation becomes more and more independent on the trajectory. It is predicted that a long-lived helical spin mode emerges~\cite{Malshukov2000,Kiselev2000}. Here we visualize this transition experimentally in a GaAs quantum-well structure with isotropic SOI. Spatially resolved measurements show the formation of a helical mode already for non-quantized and non-ballistic channels. We find a spin-lifetime enhancement that is in excellent agreement with theoretical predictions. Lateral confinement of a 2D electron gas provides an easy-to-implement technique for achieving high spin lifetimes in the presence of strong SOI for a wide range of material systems.
}

\begin{figure*}[t]
\includegraphics{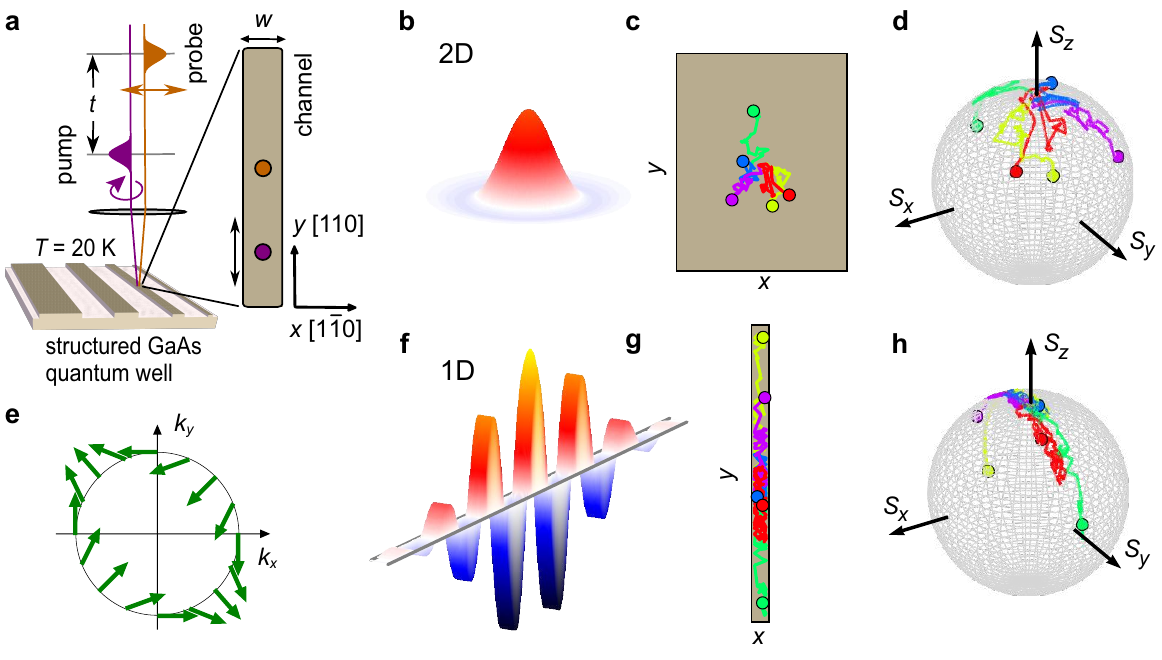}
\caption{\textbf{Measurement principle and expected spin modes. a,} A focused pump laser pulse locally excites out-of-plane spin polarization $S_z$ with a Gaussian width of $\approx$1.1$\mu$m. A probe laser pulse measures the local spin distribution after a time delay $t$ via the magneto-optical Kerr effect. Pump and probe spots are spatially scanned against each other. \textbf{b,} In the 2D case, spin diffusion in the isotropic SOI-field (\textbf{e}) turns a local spin excitation along $z$ into a Bessel-type spin mode, whose $S_z$ component is shown (red color indicates positive, blue color negative $S_z$). The spin-trajectories in the 2D plane, \textbf{c}, are correlated with the trajectories on the Bloch sphere, $\textbf{d}$, such that the local spin density decays more slowly than that of the whole ensemble. \textbf{f,} in the 1D case, the emerging spin mode is long-lived and described by a cosine oscillation of $S_z$, corresponding to a helical rotation of the spin polarization. \textbf{g,} The lateral confinement restricts the diffusive trajectories in real space, such that the spins on the Bloch sphere evolve on a ring (\textbf{h}) whose width scales with the channel width, $w$. The smaller $w$, the more the spins precess about a single axis, leading to a drastic increase of the lifetime of the helical mode.
\label{fig0} }
\end{figure*}

In a diffusive electron system with intrinsic SOI (e.g., of Rashba or Dresselhaus type), the effective spin-orbit field changes after each scattering event. This leads to a randomization of spin polarization that is described by the Dyakonov-Perel (DP) spin-dephasing mechanism~\cite{Dyakonov1972}, in the case of an initially homogenous spin excitation. Given a local spin excitation, a spin mode emerges that is described by the Green's function of the spin diffusion equation~\cite{Froltsov2001, Stanescu_2007, Liu2012}. For a 2D system in the weak SOI limit, analytical solutions exist for a few special situations, such as for the persistent spin helix case with equal Rashba and Dresselhaus SOI~\cite{Schliemann2003, Bernevig2006, Koralek2009, Kohda2012, Walser2012}. In the isotropic limit (either only Rashba or only linear Dresselhaus SOI), the spin mode is described by a Bessel-type oscillation in space (see Fig.~\ref{fig0}b)~\cite{Froltsov2001}. The spin lifetime of such a mode is only slightly enhanced~\cite{Froltsov2001} compared with the DP time because rotations about varying precession axes (see Figs. \ref{fig0}c-e) do not commute and therefore the spin polarization at a given position depends on the trajectory on which the electron reaches that position. If the electron motion is laterally confined by a channel structure of width $w$, the spin motion is restricted to a ring on the Bloch sphere (see Figs.~\ref{fig0}g and \ref{fig0}h). In this situation, the spins collectively precess along the channel direction (Fig. \ref{fig0}f)~\cite{Malshukov2000, Kiselev2000, Kettemann2007}. This extends even into the 2D diffusive regime as long as the cumulative spin rotations attributed to the lateral motion are small, i.e., as long as $w q^0 < 1$, where $q^0$ is the lateral wave number of the 2D spin mode. As a consequence, for a 2D diffusive system, increasing lateral confinement is predicted to result in an enhanced spin lifetime proportional to $(q^0 w)^2$~\cite{Malshukov2000, Kiselev2000}. This effect could be highly relevant for spintronics applications because it circumvents the conventional trade-off between a long spin lifetime and strong SOI. It has been experimentally explored in different ways, including measurements of weak-antilocalization \cite{Schaepers2006, Kunihashi2009}, the inverse spin-Hall effect \cite{Wunderlich2010}, and time-resolved Kerr rotation \cite{Holleitner2006}. None of these works were able to resolve the spin dynamics both spatially and temporally, and a quantitative investigation of the spin mode in the confined channel is still lacking.

We experimentally explore the dynamics and spatial evolution of electron spins in a 2D electron gas hosted in a symmetrically confined, 12-nm-wide GaAs/AlGaAs quantum well where the linear Dresselhaus SOI is much larger than the Rashba or the cubic Dresselhaus SOI, thus providing an almost isotropic SOI. To study the transition from 2D to 1D, we have lithographically defined wire structures along the [1$\bar{1}$0] ($x$) and [110] ($y$) directions with the channel width $w$ ranging from 0.7 to 79 $\mu$m.

%

\begin{figure*}[t!]
\includegraphics{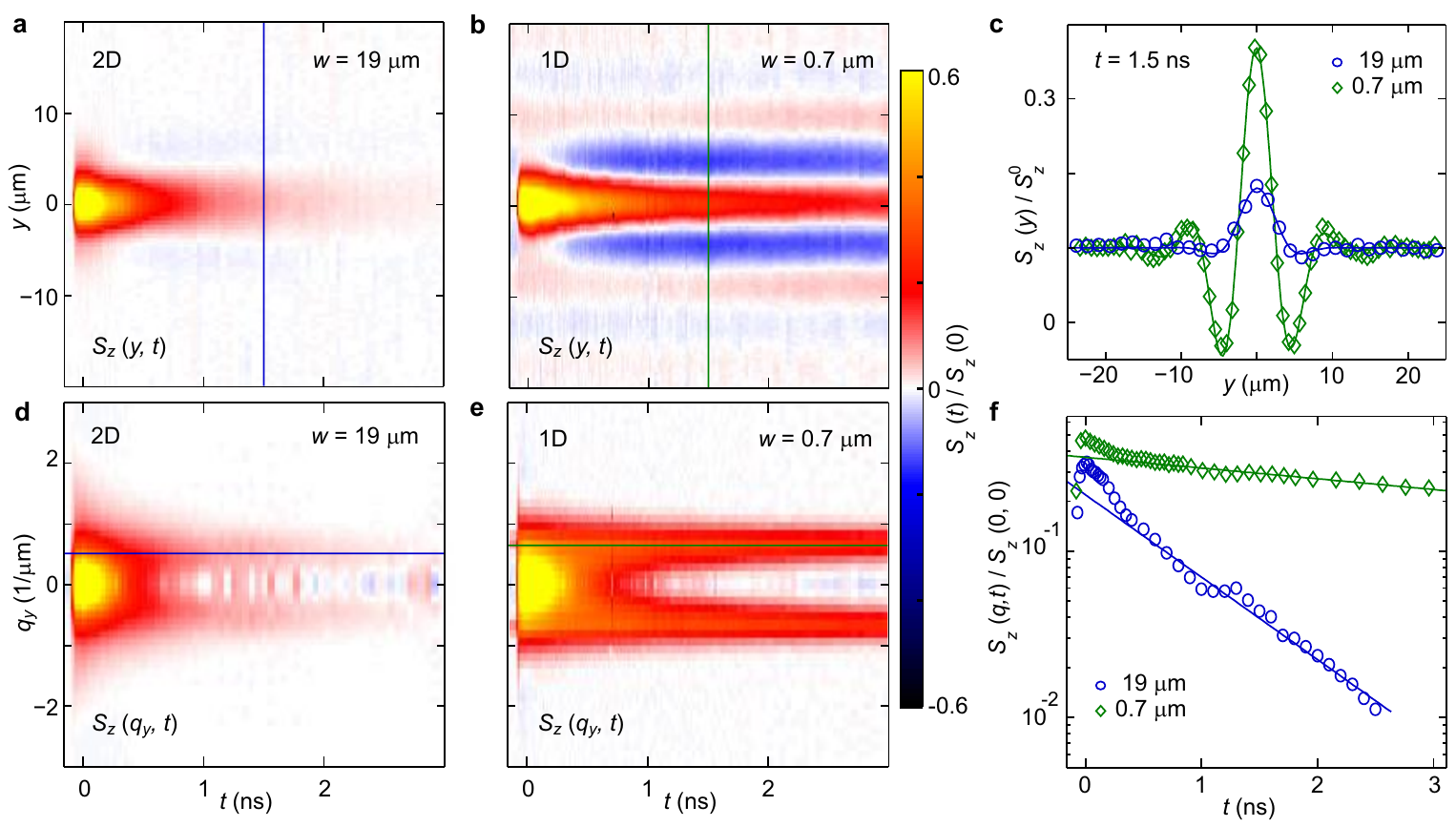}
\caption{\textbf{Direct mapping of spin precession and spin decay. a-b,} Spatial maps along the channel direction $y$ of the out-of-plane spin density $S_z$ for varying time delays $t$ between the pump and probe laser pulses. The 19-$\mu$m-wide channel (\textbf{a}) represents a 2D situation, whereas the 0.7-$\mu$m-wide channel (\textbf{b}) is close to the 1D limit and exhibits a long-lived mode with oscillating $S_z(y)$ [sign encoded as red (+) and blue (-)]. The mode from the preceding laser pulse (pulse period 12.6\,ns) is still visible at negative delay. \textbf{c,} Line-cuts through the data of (\textbf{a}) and (\textbf{b}) at $t=1.5$~ns for comparison. \textbf{d} and \textbf{e} show the Fourier transform of $S_z(y)$ for the 19- and 0.7-$\mu$m channel, respectively. For $w=0.7$\,$\mu$m, the initially Gaussian spectrum quickly converges to a long-lived spin mode at finite $q_y=q_y^0$. \textbf{f,} Line-cuts through the data of (\textbf{d}) and (\textbf{e}) at $q_y\approx q_y^0$. At each $q_y$, the amplitude decays biexponentially, with the the longer-lived mode visible at longer $t$.
\label{fig1} }
\end{figure*}

Figure \ref{fig0}a shows a sketch explaining the measurement principle.
Spins polarized along the out-of-plane direction, $z$, are locally excited at time $t=0$ by a focused, circularly polarized pump laser pulse, which has a Gaussian intensity profile of a sigma-width of 1.1~$\mu$m. A second, linearly polarized probe pulse measures the out-of-plane component, $S_z$, of the local spin density using the magneto-optical Kerr effect. The spatial evolution of the spin packet is mapped out along the channel direction for various time delays, $t$, between pump and probe pulse. All measurements have been performed at a sample temperature of 20\,K.



A measurement of spatially resolved spin dynamics in a channel in the 2D limit ($w = 19$\,$\mu$m) is shown in Fig.~\ref{fig1}a. Spins are excited at $t=0$ and at $x=y=0$ and traced as a function of $y$ and $t$. At $y = 0$, $S_z$ simply decays in time. It reverses its sign after $t > 400~$ps for electrons that diffused along $y$ by more than $\approx 4 ~\mu$m, seen as a faint blue color in the Figure. The situation is different in the $0.7$-$\mu$m-wide channel (Fig.~\ref{fig1}b). Here, spin decay is strongly suppressed and $S_z$ reverses its sign multiple times along $y$ at later times. Note that the pattern is overlaid with the spin texture that survived from the previous pump pulse at $t = -12.6$~ns. Figure~\ref{fig1}c shows measured data of $S_z(y)$ for the 19-$\mu$m and the 0.7-$\mu$m-wide channels taken at $t = 1.5$\,ns. The comparison of the two curves clearly shows an enhanced $S_z$ and strong oscillations along $y$ in the narrow channel. This indicates a helical spin mode in the 1D case. The helical nature is further supported by measured maps where an external magnetic field is applied along the $x$ direction, rotating the helix as a function of time, see supplementary information.

For a deeper analysis, it is advantageous to Fourier-transform $S_z (x, y, t)$ into momentum space. Thereby one obtains Fourier components $S_z (q_x, q_y, t)$ at wave numbers $q_x$ and $q_y$ that according to theory decay biexponentially in time~\cite{Stanescu_2007}. For channels narrower than 15\,$\mu$m, the spin modes exhibit a pronounced structure only along the channel direction, and we therefore analyze the 1D Fourier transformation along this direction. For wider channels, we obtain the 2D Fourier transformation from 1D scans of $S_z$ by assuming a radially symmetric spin mode, see supplementary information for details. This is justified because we observe a similar dependence of $S_z$ along the $x$ and $y$ directions, as seen from the values obtained for wavenumbers $q_x^0$ and $q_y^0$ later in the text.

\begin{figure*}[t!]
\includegraphics{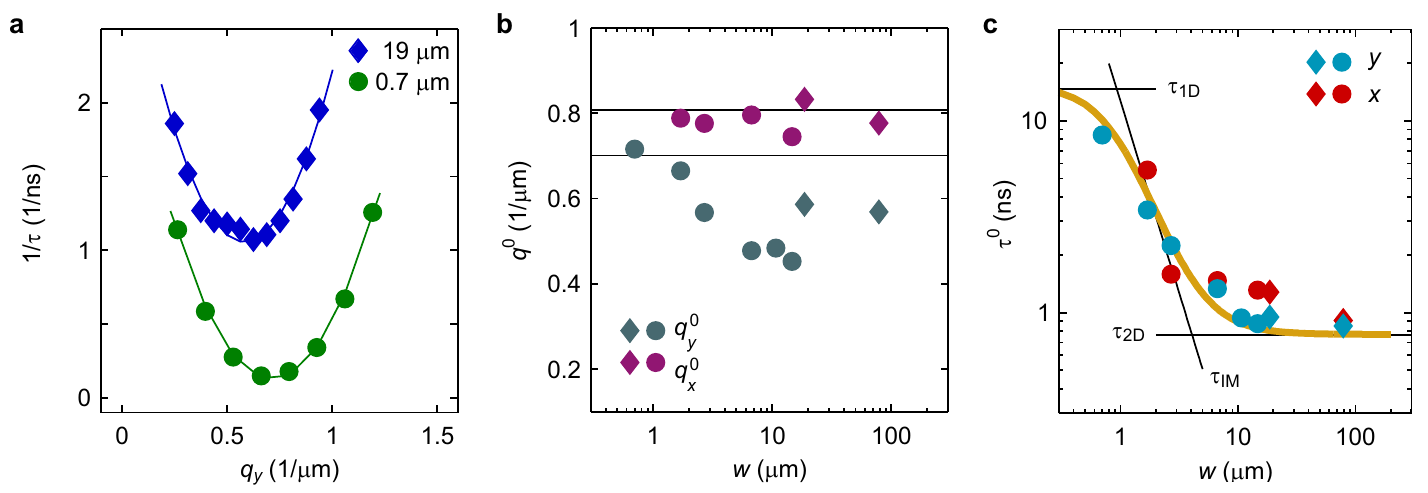}
\caption{\textbf{Fit results. a,}  Decay rates for the 19-$\mu$m (diamonds) and 0.7-$\mu$m-wide (dots) channels along the $y$ direction as a function of the wave number, $q_y$. Data is obtained from a fit of the Fourier-transformed $S_z(y)$ to an exponential decay for $t>500$\,ps, thus corresponding to the long-lived mode. Solid lines are parabolic fits of the decay rate versus $q$, from which the position, $q_y^0$, and the lifetime, $\tau^0$, of the evolving spin mode are obtained. \textbf{b,} Values for the spin-mode wave numbers $q_x^0$ and $q_y^0$, shown for measurements at various channel widths, $w$, along the $x$ and $y$ direction. \textbf{c,} Lifetime of spin modes, $\tau^0$, as a function of $w$ and for both channel directions. Solid black lines are the theoretically expected lifetimes. The yellow solid line is their interpolation. In \textbf{b} and \textbf{c}, circles (diamonds) stand for fit values obtained from 1D (2D) Fourier transformations.
\label{fig2} }
\end{figure*}

Figures \ref{fig1}d and \ref{fig1}e show $S_z (q_y, t)$ for the 19- and $0.7-\mu$m wires, respectively. The Gaussian distribution of $S_z (q_y,t)$ decays in time with very different rates for varying $q_y$, which are minimal at a finite wavenumber, $q^0_y$. Figure~\ref{fig1}f shows traces at $q_y \approx q^0_y$ for the two cases. For $t > 500$\,ps, we fit each trace with a single exponential decay to obtain the momentum-dependent lifetime $\tau (q_y)$ of the longer-lived spin mode~\cite{Koralek2009, Stanescu_2007}. The decay rates, $1/\tau (q_y)$ are shown in Fig.~\ref{fig2}a. In both the 1D and the 2D case, $1/\tau$ vs $q_y$ can be well approximated close to $q_y^0$ by the parabolic function~\cite{Liu2012, Stanescu_2007}

\begin{equation}
1/\tau = 1/\tau^0 + D_s (q_y - q_y^0)^2,
\label{parabola}
\end{equation}

where $D_s$ is the spin diffusion constant \footnote{Note that the spin diffusion constant differs from the electron diffusion constant measured by transport measurements because it is sensitive to electron-electron scattering.}. Figures \ref{fig2}b and \ref{fig2}c plot the values obtained for $q_y^0$ ($q_x^0$) and $\tau^0$, respectively, for channels along the $y$ ($x$) direction and of various widths.

\begin{figure}[b]
\includegraphics{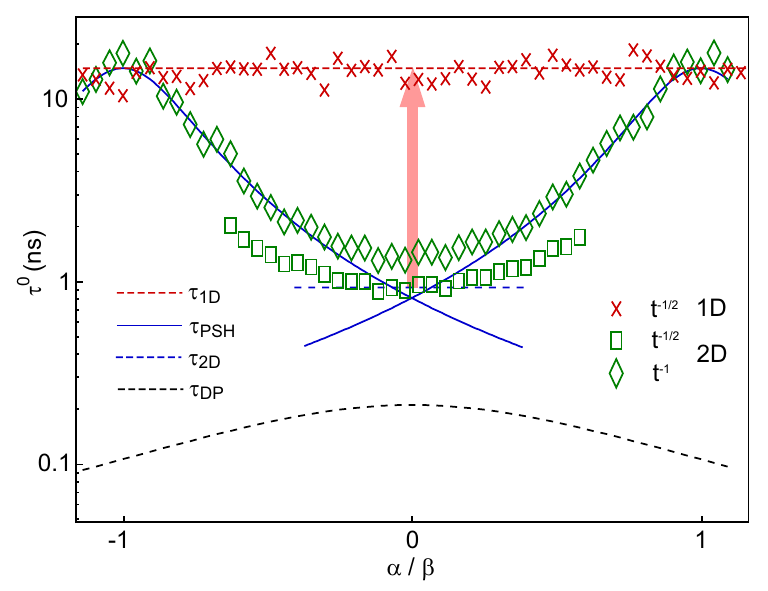}
\caption{\textbf{Lifetime enhancement for various $\alpha / \beta$.} Lifetimes for the 1D and 2D situation as determined from Monte-Carlo simulations for various ratios of $-1.1 < \alpha / \beta < 1.1$. Data is obtained by fitting $S_z(y,t)$ with a model that includes a diffusive dilution proportional to either $1/t$ or $1/\sqrt t$. The former is used in the 2D case for $|\alpha|\approx|\beta|$ (diamonds), the latter for isotropic SOI in the 2D case (rectangles) and for the 1D case (crosses). The solid and dashed lines are theoretical curves (see supplementary information). The lifetime enhancement under lateral confinement is largest for $\alpha = 0$ (arrow). For both the 2D case at $|\alpha| = |\beta|$ and the 1D case, the lifetime is limited by the same value given by the cubic SOI only.
\label{fig3} }
\end{figure}

Comparing $q_x^0$ and $q_y^0$ in Fig.~\ref{fig2}b, we observe a slight anisotropy characterized by $q_x^0 > q_y^0$. This means that the SOI is stronger for electrons that move along $x$ and indicates a remaining Rashba field due to a slight asymmetry in the quantum well. The SOI coefficients, $\alpha$ and $\beta$, are obtained from $q_{y}^0$ and $q_{x}^0$ measured in the 1D limit by using the expressions

\begin{eqnarray}
q_\textrm{y}^0 = \Big| \frac{2 m^*}{\hbar^2} \left( \alpha - \beta \right) \Big| \approx 0.7~\mu \textrm{m}^{-1} ,~\textrm{and}\\
\: q_\textrm{x}^0 =  \Big| \frac{2 m^*}{\hbar^2} \left( \alpha + \beta \right) \Big| \approx 0.8~\mu \textrm{m}^{-1}.
\label{q0s}
\end{eqnarray}

Here, $m^*$ is the effective electron mass and $\hbar$ is the reduced Planck's constant, $\alpha$ is the SOI parameter of the Rashba field and $\beta = \beta_1 - \beta_3$ that of the Dresselhaus field. $\beta_1$ and $\beta_3$ characterize the linear and cubic Dresselhaus fields, respectively. Values for $\alpha$, $\beta_1$, $\beta_3$ and $D_s$ are given in the supplementary information.

The dependence of $q_x^0$ on $w$ is rather flat, whereas $q_y^0$ decreases for increasing $w$. This is in agreement with the prediction that $q_y^0$ of the 2D spin mode is smaller for slightly anisotropic SOI than expected from Eq.~(\ref{q0s})~\cite{Stanescu_2007, Poshakinskiy2015}. Close to the persistent spin helix situation, the same effect leads to a suppression of precession along $y$.

The lifetime, $\tau^0$, however, behaves almost identically for both wire directions and increases by about one order of magnitude from $w = 19$ to $0.7~\mu$m.
Theory provides expressions for the lifetime in the 2D limit \cite{Froltsov2001, Stanescu_2007}, $\tau_\textrm{2D}$,  and in the intermediate regime \cite{Kiselev2000, Malshukov2000}, $\tau_\textrm{IM}$, see supplementary information. For very narrow channels, the lifetime $\tau_\textrm{1D}$ is limited by cubic Dresselhaus SOI only, and as we will show later, is the same as in the completely balanced spin-helix case~\cite{Salis2014}.
The theoretically expected values are plotted in Fig.~\ref{fig2}c as black lines.
The interpolation between $\tau_\textrm{2D}$, $\tau_\textrm{IM}$ and $\tau_\textrm{1D}$ (yellow line in Fig.~\ref{fig2}c) is in very good quantitative agreement with the experimental data. Although $\tau^0$ towards smaller $w$ is not yet saturated, it is possible to project that cubic SOI will limit the lifetime.


The lifetime enhancement achievable by channel confinement depends strongly on the ratio $\alpha / \beta$. Figure \ref{fig3} shows lifetimes determined by Monte-Carlo simulations for $-1.1 < \alpha / \beta < 1.1$. For this analysis, we determined $\tau^0$ directly from the decay of $S_z (y=0, t)$, see supplementary information. Without Fourier transformation one has to account for a diffusion factor that reduces the amplitude, in addition to an exponential decay term. The diffusive dilution of electrons in 2D scales with $1/t$ and in 1D with $1/\sqrt{t}$. Interestingly, the spins in a 2D system, however, also decay with $1/\sqrt{t}$ for the isotropic SOI case~\cite{Stanescu_2007}. Solid and dashed lines are the theoretically expected values of $\tau^0$ for 2D and 1D spin modes ($\tau_\textrm{1D}$, $\tau_\textrm{2D}$, $\tau_\textrm{PSH}$), as well as for the DP case ($\tau_\textrm{DP}$).

We find that in a narrow channel, $\tau^0$ does not depend on $\alpha$ or $\beta_1$ and is limited by cubic SOI ($\beta_3$) only. The same limit is reached in the 2D situation at $\alpha = \beta$, i.e., when the system is tuned to the persistent spin helix symmetry. For given SOI coefficients, the maximal lifetime enhancement under lateral confinement in the diffusive limit occurs for the isotropic case ($\alpha=0$). Close to $\alpha = \beta$, the lifetime enhancement is small, but a reduction of diffusive dilution was observed~\cite{Altmann2014}.


In conclusion, we measured the evolution of a local spin excitation in a GaAs/AlGaAs quantum well dominated by linear Dresselhaus SOI. Because of SOI, the lateral confinement leads to an increased correlation between electron position and spin precession. Using a real-space mapping of the spin distribution, we observe a helical spin mode accompanied by an enhanced lifetime for decreasing channel width. The analysis in momentum space shows that the long-lived components decay exponentially with a minimum rate at a finite $q^0$.
Both the precession length and the lifetime are in quantitative agreement with theory for the 2D limit, the 1D limit and also for the intermediate regime. 
The narrowest channel in our study still is 10 times wider than the mean-free-path (including electron-electron scattering) and 100 times wider than the Fermi wavelength of the electrons. At those smaller length scales, also a reduction of the cubic SOI contribution to spin decay was predicted~\cite{Wenk2011}.

These findings illuminate an interesting path for studying spin-related phenomena. Lateral confinement provides a straight forward method for achieving spin lifetimes that are otherwise only possible by careful tuning of SOI to the persistent spin helix symmetry. This facilitates the use of spins in materials with stronger SOI, such as InAs or GaSb, but also in group-IV semiconductors, like Si and Ge. Extending the presented method to 1D systems in the quantized limit will be relevant for the quest for Majorana fermions when combined with superconductors~\cite{Lutchyn2010, Oreg2010, Alicea2010, Mourik2012}. Furthermore, the results are important for transport studies and transistor applications~\cite{Schliemann2003, Kunihashi2012APL, Chuang2014} using SOI in 1D or quasi-1D systems.

We acknowledge financial support from the NCCR QSIT and from  the Ministry of Education, Culture, Sports, Science, and Technology (MEXT) in Grant-in-Aid for Scientific Research Nos. 15H02099 and 25220604.
We thank R. Allenspach, A. Fuhrer, T. Henn, A.~V.~Poshakinskiy, F. Valmorra, M. Walser, and R.~J. Warburton for helpful discussions, and U. Drechsler for technical assistance.

\newpage

\section{Supplement}

\textbf{Measurement Setup}

We use a time- and spatially-resolved pump-probe technique to map out spin dynamics. Two Ti:sapphire lasers are used to generate the pump and probe laser pulses at 785~nm and 802~nm, respectively. Pulse lengths are on the order of 1 ps and the repetition rate is 79.1~MHz, which corresponds to 12.6~ns between two pulses. The pulses of the two lasers are electronically synchronized. The delay between pump and probe pulses is controlled by a mechanical delay line. The linearly polarized probe beam is chopped at a frequency of 186~Hz. The polarization of the pump is modulated by a photo-elastic modulator between plus and minus circular polarization at a frequency of 50~kHz. The sample is located inside an optical cryostat at a temperature of 20~K. Both laser beams are focused onto the sample surface with a lens inside the cryostat. The Gaussian sigma width of the intensity profile is $\approx 1.1~\mu$m for both spots. The power of the pump beam is $100~\mu$W and that of the probe beam is $50~\mu$W. The pump spot is positioned onto the sample relative to the fixed probe spot by a scanning mirror. After being reflected from the sample, the pump beam is blocked with a suitable edge filter, whereas the probe beam is sent to a detection line, where its polarization is monitored by a balanced photodiode bridge and lock-in amplifiers.

\textbf{Sample preparation}

A GaAs quantum well is grown on a (001) GaAs substrate by molecular beam epitaxy. The barrier material is Al$_{0.3}$Ga$_{0.7}$As. Front and back Si $\delta$-doping layers are positioned such that the electric field perpendicular to the quantum-well plane is very small. A sheet density of $3.5 \times 10^{15}$~m$^{-2}$ and a transport mobility of $7.0 \times 10^5$~cm$^2$(Vs)$^{-1}$ were determined at 4~K after illumination by a van-der-Pauw measurement. A $5 \times 5$~mm$^2$ piece was cleaved out of the 2'' wafer and processed with photo-lithography. Wires of variable width were etched by wet-chemical etching. The effective widths of the wires as given in the main text were determined by scanning electron microscopy images, measuring the width of the top surface.

\textbf{Theory}

The measured values of $q_y^0$ and $q_x^0$ in the 1D limit allows the determination of $\alpha$ and $\beta$, as described in the main text. Additionally, the knowledge of the electron density, $n_s$, allows the calculation of the cubic Dresselhaus coefficient via $\beta_3 = -\gamma \times n_s /4$ with $\gamma =-11 \times 10^{-30}$~eVm$^3$~\cite{Walser2012PRB}. Equation (\ref{parabola}) allows the determination of $D_s$. The structure investigated is described by the following parameters.

\begin{center}
\begin{tabular}{|c|c|c|c|}
$D_s$ & $\alpha $ & $\beta_1$ & $\beta_3$ \\ \hline
0.005 m$^2$/s & -0.3 meV\r{A} & 4.9 meV\r{A} & 0.6 meV\r{A}
\end{tabular}
\end{center}

The spin-dephasing time in the 2D limit for Dresselhaus fields only is given by \cite{Stanescu_2007}

\begin{align}
\begin{split}
\tau_\textrm{2D} = \Bigg[ 2 \frac{D_s m^{*2}}{\hbar^4} \Bigg( \beta^2 + 3 \beta_3^2 - 2\beta^2   - \left( \frac{\beta^2 + \beta_3^2}{8 \beta^2} \right) \Bigg) \Bigg]^{-1} ~.
\end{split}
\label{tau2D}
\end{align}

The spin-dephasing time in the 2D limit close to the persistent spin helix symmetry is given by \cite{Bernevig2006, Salis2014}

\begin{equation}
\tau_{\textrm{PSH}} = 2 D_s \frac{m^{*2}}{\hbar^4} \left[ (\alpha - \beta)^2 + 3 \beta_3^2 \right] \,.
\label{tauPSH}
\end{equation}

The spin-dephasing time in the 1D limit  is given by \cite{Chang2009, Wenk2011, Salis2014}

\begin{equation}
\tau_\textrm{1D} = \left[ 6 \frac{D_s m^{*2}}{\hbar^4} \beta_3^2  \right]^{-1} ~.
\end{equation}

The Dyakonov-Perel spin-dephasing time for out-of-plane spin polarization is given by

\begin{equation}
\tau_\textrm{DP} = \left[ 8 \frac{D_s m^{*2}}{\hbar^4} \left( \alpha^2 + \beta^2 + \beta_3^2 \right) \right]^{-1} ~.
\label{tauDP}
\end{equation}

The following scaling behavior is expected for the intermediate regime between the 1D and the 2D limit \cite{Chang2009}:

\begin{equation}
\tau_\textrm{IM} = 48 \tau_\textrm{DP} \left( q_0 w \right)^{-2} ~.
\end{equation}

\textbf{Real-space evaluation}

\begin{figure}
\includegraphics{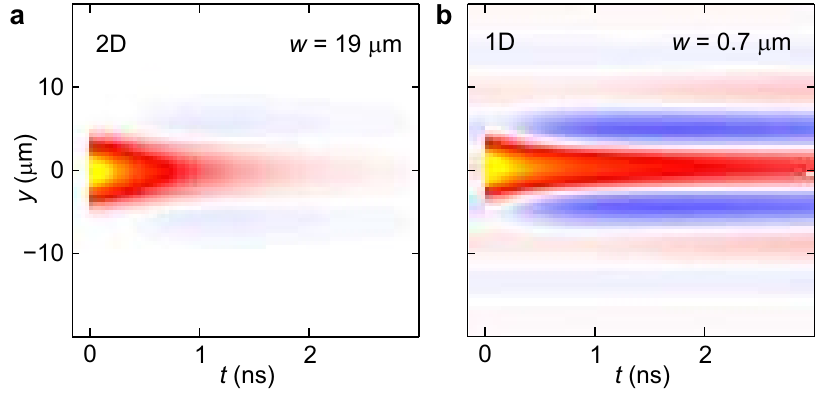}
\caption{\textbf{Fits of $S_z (y, t)$. a,} Fit of a 2D real space model, as described in the text, to the data of the $19-\mu$m-channel. \textbf{b,} Fit of a 1D real space model, as described in the text, to the data of the $0.7-\mu$m-channel. The values determined by this method for $\tau^0$ agree with the evaluation of $S_z (q_y, t)$ in momentum-space.
\label{figSupReal} }
\end{figure}

An analysis of the spin dynamics is, in principle, also possible in the real-space representation $S_z (y, t)$ and a model can be fitted to the full data. It contains the spatial variation of the spin mode (Bessel function in 2D, cosine in 1D), a Gaussian envelope, which originates in the Gaussian shape of the initial spin density profile, the diffusive broadening of this Gaussian envelope, and an exponential decay~\cite{Altmann2014}. A diffusive decay term accounts for the diffusive dilution of the spin density. As mentioned in the main text when discussing the evaluation of $\tau^0$ from Monte-Carlo simulations, this term is proportional to $1/\sqrt{t}$ in the 1D situation. In the 2D case, however, it is either $1/t$ for $\alpha \approx \beta$ or $1/\sqrt{t}$ for the isotropic case, i.e., when $\alpha \beta = 0$ \cite{Stanescu_2007}. An additional complication arises because the initial spin excitation has a finite spatial distribution owing to the pump-laser spot size. This can be accounted for by a convolution of the exact solution for a $\delta$-peak excitation with a Gaussian function that is itself a convolution of the intensity profile of both the pump and the probe laser spots. Figure \ref{figSupReal} shows such fits for the 19-$\mu$m and 0.7-$\mu$m wires along the $y$ direction. This procedure is numerically more demanding than the fits of the exponentially decaying Fourier components of $S_z(y)$, and the fit parameter $\tau^0$ is obtained more indirectly. We therefore prefer to evaluate the Fourier-transformed $S_z$.

\textbf{Fourier transformation}

To obtain the decay dynamics at specific wave-vectors, the spatial spin pattern $S_z(x,y)$ needs to be Fourier-transformed.
If we consider a 1D system defined by a channel along $y$, $S_z$ only varies along $y$, which allows us to apply a 1D Fourier transformation:

\begin{equation}
S_z (q_y,t) = \int_{- \infty}^\infty  \cos (q_y y) S_z (y, t) \mathrm{d}y ~.
\end{equation}

In the 2D situation, the Fourier transformation in principle requires full knowledge of $S_z(x,y)$. Because of the almost isotropic SOI, we can assume a radially symmetric mode and obtain the Fourier component $S_z(q_y,t)$ from scans of $S_z$ along $y$:

\begin{equation}
S_z (q_y,t) = \int_0^\infty 2 \pi y \mathrm{J}_0 (q_y y) S_z (y, t) \mathrm{d}y ~.
\end{equation}

\begin{figure}
\includegraphics{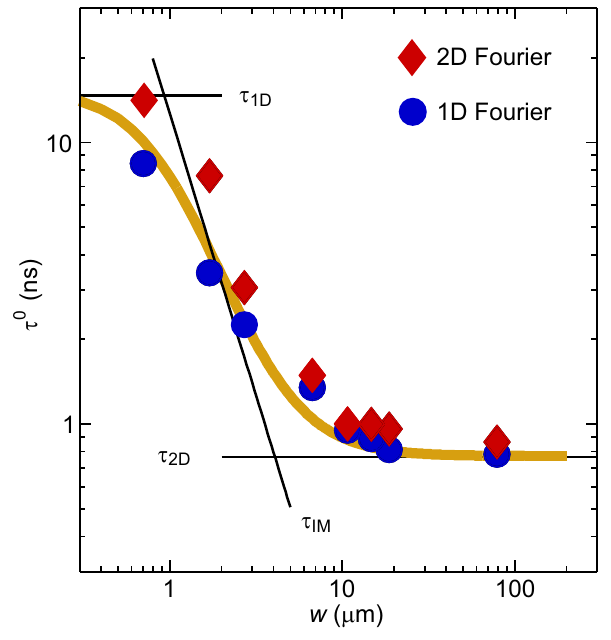}
\caption{\textbf{Comparison of radial and linear Fourier transformation} Experimental values of $\tau^0$ as obtained by 2D (diamonds) and 1D (circles) Fourier transformations of $S_z (y, t)$. Also shown are the theoretical values of $\tau_1D$, $\tau_2D$ and $\tau_\textrm{IM}$, as well as their interpolation.
\label{figSup1} }
\end{figure}

Here, $\mathrm{J}_0$ is the zeroth-order Bessel function. When the channel width $w$ is gradually reduced, the system undergoes a transition from the 2D to the 1D situation. We have Fourier-transformed all data sets with both methods and fitted them as described in the main text. Figure \ref{figSup1} shows $\tau^0$ along the $y$-direction determined with both methods. 
The values are very similar. In the main text, we therefore plot the 2D transformation for $w \geq 15~\mu$m and the 1D transformation for narrower channels.

\textbf{The sign of $\alpha$}

\begin{figure}
\includegraphics{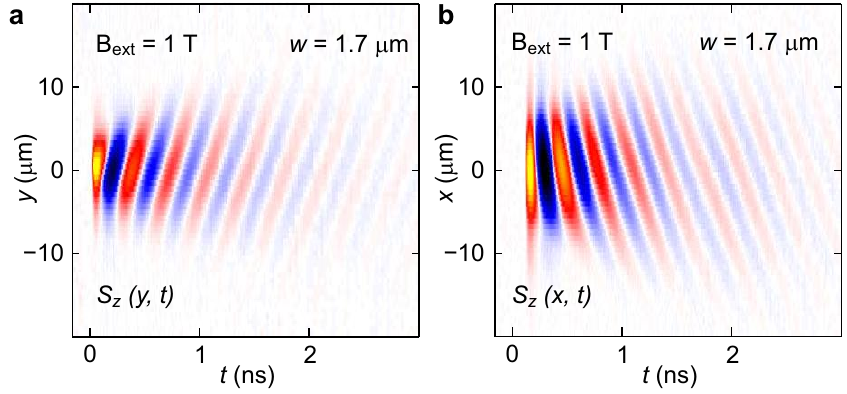}
\caption{\textbf{Spin maps at an external magnetic field. a,} $S_z (y, t)$ in a 1.7-$\mu$m-wide wire along the $y$ direction at an external magnetic field of $B_\textrm{ext} = + 1 $~T. The field is perpendicular to the wire direction. The sample temperature is 30~K. \textbf{b,}  $S_z (x, t)$ in a 1.7-$\mu$m-wide wire along the $y$ direction at an external magnetic field of $B_\textrm{ext} = + 1 $~T. The field is perpendicular to the wire direction. The sample temperature is 10~K. The magnetic field induced additional spin precession lines with constant phase are tilted, showing the helical nature of the spin mode. From the opposite sign of the tilts in the $(x, t)$ and the $(y, t)$ planes, it is concluded that $\alpha$ is negative.
\label{figSupTilt} }
\end{figure}

While $\beta$ can only be positive, the sign of $\alpha$ depends on the direction of the perpendicular electric field with respect to the growth direction. To determine the sign of $\alpha$, we measure spatial spin maps also at an applied external magnetic field. Figure \ref{figSupTilt} shows such measurements in a channel with $w = 1.7~\mu$m for an external magnetic field of $B_\textrm{ext} = +1$~T perpendicular to the wire direction. Evaluating these measurements as done in \cite{Walser2012}, we can conclude that $\alpha < 0$. 
Moreover, the continuous lines of constant spin orientation in these measurements demonstrate the helical nature of the ground mode. 

\textbf{Monte-Carlo simulations}

Spin dynamics in a laterally confined 2D electron gas are calculated numerically using a Monte-Carlo method where the positions and spin orientations of $3 \times 10^5$ electrons are updated in time steps of 0.1 ps. Electrons are distributed on a Fermi circle and scatter isotropically, with the mean scattering time given by $\tau=2 D/v_F^2$, where $v_F=\hbar k_F/m$ is the Fermi velocity. Each electron moves with the Fermi velocity and sees an individual spin-orbit field as defined in the supplementary information of Ref.~\cite{Walser2012} that depends on its velocity direction. The real-space coordinates and the corresponding spin dynamics are calculated semiclassically. We initialize the electrons at $t=0$ all with their spins oriented along the $z$ direction and distribute their coordinates in a Gaussian probability distribution with a center at $x=y=0$ and a $\sigma$ width of 0.5\,$\mu$m. Histograms of the electron density and the spin orientations are recorded every 5\,ps, and the simulation is run until $t=5$\,ns is reached. We obtain the spin polarization at $x=y=0$ versus $t$ from the spin-density maps using a convolution with an assumed Gaussian probe spot size of 0.5\,$\mu$m. We determine the spin lifetimes $\tau^0$ by fitting the transients with a function proportional to $1/t\times\exp -t/\tau^0$ or $1/\sqrt{t} \times \exp -t/\tau^0$ in a window 800\,ps~$<t<$\,4000\,ps, where additional spin decay is negligible because of the small spot sizes \cite{Salis2014}. For the data shown in Fig. 4, we have used the following parameters: $D_s=0.004$ m$^2$/s, $n_s=3.4\times$10$^{15}$\,cm$^{-2}$, $\beta_1 = 4.9\times10^{-13}$\,eVm and $\beta_3=0.6\times10^{-13}$\,eVm. Lateral confinement was implemented by assuming specular scattering at the channel edges. For the 1D case, $w=0.4$\,$\mu$m was used.

%

\end{document}